\documentclass[pra,twocolumn,aps]{revtex4}
\usepackage{graphicx}  
\usepackage{dcolumn}   
\usepackage{bm}        
\usepackage{amssymb}   
\usepackage{setspace} 

\begin{document}

\title{Power-Broadening-Free Correlation Spectroscopy in Cold Atoms}

\author{H. M. Florez $^{1}$, L.S. Cruz$^{2}$, M. H. G. de Miranda$^{1,3}$, R.A. de Oliveira$^{3}$,  J.W.R. Tabosa$^{3}$,  M. Martinelli$^{1}$, and D. Felinto$^{3}$}

\affiliation{$^{1}$Instituto de F\'{\i}sica, Universidade de S\~ao Paulo, 05315-970 S\~ao Paulo, SP-Brazil \\ $^{2}$Universidade Federal do ABC, Santo Andr\'e, SP-Brazil \\ $^{3}$Departamento de F\'{\i}sica, Universidade Federal de Pernambuco, 50670-901 Recife, PE-Brazil}

\begin{abstract} 
We report a detailed investigation on the properties of correlation spectra for cold atoms under the condition of Electromagnetically Induced Transparency (EIT). We describe the transition in the system from correlation to anti-correlation as the intensity of the fields increases. Such transition occurs for laser frequencies around the EIT resonance, which is characterized by a correlation peak. The transition point between correlation and anti-correlation is independent of power broadening and provides directly the ground-state coherence time. We introduce a method to extract in real time the correlation spectra of the system. The experiments were done in two distinct magneto-optical traps (MOT), one for cesium and the other for rubidium atoms, employing different detection schemes. A simplified theory  is introduced assuming three-level atoms in $\Lambda$ configuration interacting with a laser with stochastic phase fluctuations, providing a good agreement with the experimental observations.
\end{abstract}

\maketitle

\section{introduction}

Advances in quantum optics have increased the attention to protocols using light as a quantum information carrier~\cite{Bennett84,Duan2001,knill01}. Information can be encoded in different degrees of freedom of the electromagnetic field, such as  amplitude and phase. In order to encode and process information,  a  medium is necessary to intermediate the interaction between two (or more) electromagnetic fields. Ensembles of neutral atoms are good candidates as such medium because they can provide a set of coherent interactions to manipulate the properties of light in a controllable way~\cite{Lukin03}. Among these coherent interactions, Electromagnetically Induced Transparency  (EIT)~\cite{Harris,HarrisExp,Marangos05,Novikova11} is widely employed to modify the absorptive and dispersive properties of an atomic medium. On the other hand, the interaction between fields and atoms also modifies the properties of light. These modifications can be measured by a technique called Noise Spectroscopy. This high-resolution spectroscopic technique was introduced by Yabusaki and collaborators~\cite{Yabusaki91} to study the D1 and D2 absorption spectra of rubidium and  cesium atoms.

Since its introduction, there has been a large number of experiments on noise spectroscopy in vapor cells. Among others groups, we employed this technique to observe correlations and anti-correlations in the fluctuations of two laser beams after interacting with an atomic vapor at room temperature fulfilling the conditions for EIT~\cite{Garrido03,Martinelli04,Cruz07, Scully05,Scully08,Scully10,Xiao09}. However, the observations in this kind of sample always suffer from Doppler broadening, which mixes the contribution of atoms from different velocity classes, or even different transitions, in the detected signal. As a consequence, the interpretation of the observed signals becomes considerably more complicated~\cite{Cruz07}.

A clearer picture of the process, with a straightforward comparison between theory and experiment, is obtained by using cold atomic ensembles for such correlation measurements, since these systems have virtually no Doppler broadening and the different transitions can be easily isolated from each other. In this work, we investigate a technique for correlation spectroscopy which is power broadening free. We perform an in-depth study of such correlation spectra in cold samples of two different alkali atoms, rubidium and cesium, as a function of detuning and intensity of the fields. The cold samples are obtained from Magneto-Optical Traps (MOT). This mapping provides a detailed description of the dynamics in frequency space of both correlations and anti-correlations for EIT-like atomic systems.

As previously observed in vapor cells \cite{Scully05,Xiao09}, an interesting feature of these systems is that the width of the EIT correlation peak does not depend on the intensity of the exciting fields. In Ref.~\cite{Felinto13} we found a simple physical interpetration for this and other crucial features of the problem, where we explored only same-time correlations between the two EIT fields. This temporal approach has clear advantages for the intuitive understanding of the signal, but it is  susceptible to spurious electronic noises over a large band of frequencies. A much better signal-to-noise ratio may be obtained by working in frequency space, selecting a single analysis frequency for which we have particularly low electronic noise. For this reason, the systematic experimental and theoretical investigation reported here is all conducted in such frequency space.

The paper is organized as follows: In Sec.~\ref{sec:experiment},  we present our setups and experimental results, with two subsections describing the similarities and differences of the setups for the two atomic species. These two setups have quite different detection schemes and provide complementary experimental approaches to the problem. In Sec.~\ref{sec:cesiummot}, the results for the cesium MOT are used as a bridge from the usual same-time-correlation approach to the analysis of the system in frequency space for a particular analysis frequency. We point out the advantages of this second method in terms of the signal-to-noise ratio. In Sec.~\ref{sec:rubidiummot}, we provide the results for the rubidium MOT with a detection scheme designed specifically for the frequency-space approach. With this scheme, it is possible to obtain real time measurements of the correlation coefficient. In Sec.~\ref{sec:theory}, we introduce our theoretical model, which is based on a simple closed three level system excited by two fields originated from the same laser. We consider only stochastic fluctuation of the laser  coming from the phase noise associated to its linewidth. Conclusions are presented on Sec.~\ref{sec:conclusions}.

\section{Experiments}
\label{sec:experiment}

\begin{figure}[ht]
\centering
\includegraphics[width=82mm]{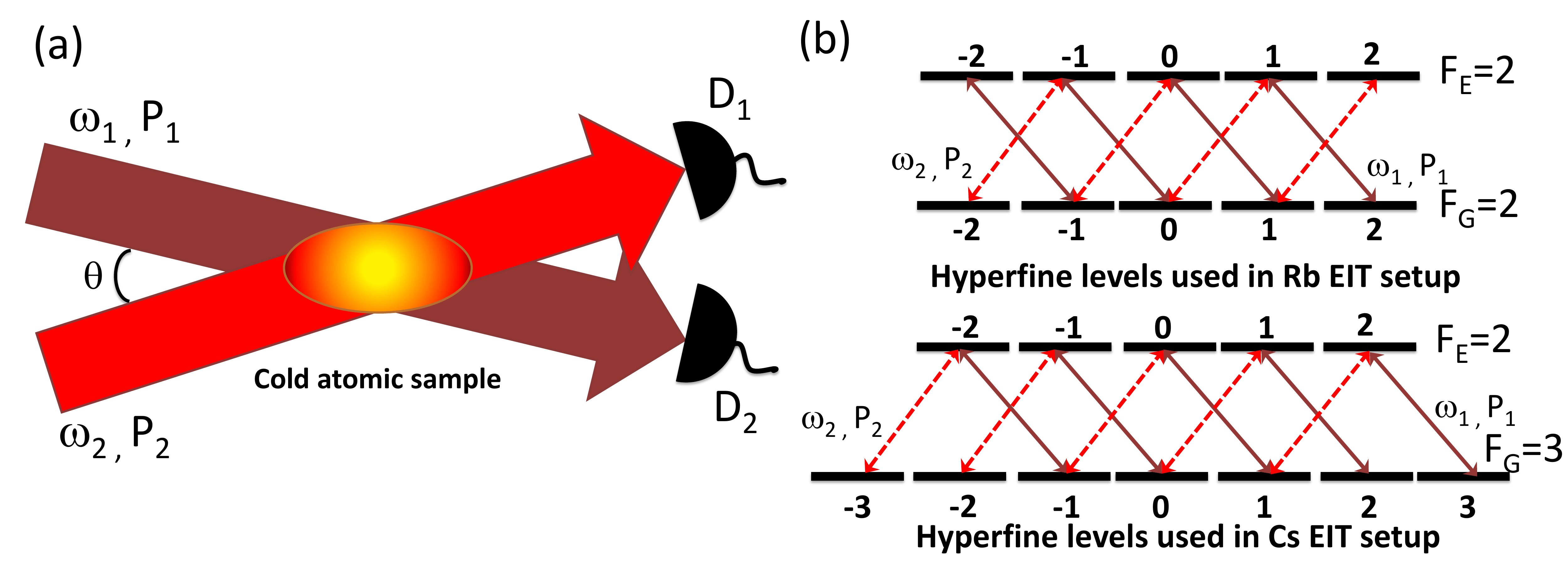}
\vspace{-0.0cm}
\caption{(a) Basic scheme for noise spectroscopy of cold atoms by optical fields under the EIT regime. The atomic sample interacts with two  electromagnetic fields with polarizations $\sigma_{+}$ and $\sigma_{-}$, respectively. After interaction with the medium, the beams are detected by photodiodes $D_1$ and $D_2$. (b) [upper] Transition $F_G=2 \rightarrow F_E=2$ used in the Rubidium experiment and [below] transition $F_G=3 \rightarrow F_E=2$ used for Cesium. We also show the beams 1 and 2 addressing the various Zeeman transitions.}
\label{fig:atom_field}
\end{figure}

In this section, we describe our series of experimental results, performed in two different MOTs, one for cesium and another for rubidium atoms. In both cases,  a cold atomic cloud interacts with two beams, $1$ and $2$, with a small angle $\theta$ between their propagation directions (see Fig.~\ref{fig:atom_field}a). The two beams come from the same External-Cavity Diode Laser (ECDL), but they have orthogonal circular polarizations ($\sigma^-$ for $1$ and $\sigma^+$ for $2$) and their frequencies $\omega_1$ and $\omega_2$ are independently tuned by distinct Acousto-optic modulator (AOM). In this way, we typically keep the $\omega_2$ fixed and tune the  $\omega_1$ across the two-photon resonance. We denote the detunings of beams $1$ and $2$ from the atomic one-photon resonance (at $\omega_{GE} = \omega_E - \omega_G$) by $\delta_1$ and $\delta_2$, respectively. The beams had similar diameters and the power was adjusted to $P_{1}$ and $P_{2}$, respectively.After transmission through the atomic ensemble, the beams are directed to two detectors, $D_1$ and $D_2$, whose signals are later analysed for their correlations.  
  
\vspace{-0.0cm}  

The relevant atomic structure in both apparatuses consists of a Degenerate Two Level System (DTLS), formed by the ground state $G$ with angular momentum quantum number $F_{G}$ and an excited state $E$ with angular momentum quantum number $F_{E}$. For null magnetic fields, the $2F_{G} + 1$ and $2F_{E} + 1$ Zeeman levels are degenerated in energy. 
Ground and excited state energies are, respectively,  $\hbar \omega_{G}$ and $\hbar \omega_{E}$. The choice of the angular momentum of the ground and excited  states establishes if the observed phenomena corresponds to a reduction of absorption (EIT) or an enhancement, a coherent effect called Electromagnetically Induced Absorption (EIA)~\cite{Arturo99a}. For observation of EIT in DTLS it is necessary that the levels obey the relations $F_{G} \geq F_{E}$ and $F_{G} \geq 1$.  In our specific case, we studied the transition $F_G=2 \rightarrow F_E=2$ for $^{85}$Rb and the transition $F_G=3 \rightarrow  F_E=2$ for $^{133}$Cs, as shown in Fig.~\ref{fig:atom_field}b.

Typically, optical fields deriving from an ECDL present only phase noise, resulting in phase correlations between fields $1$ and $2$. Without propagation through the atomic ensembles, these phase correlations are not detected in the electrical currents generated at $D_1$ and $D_2$, since the intensity measurements at the detectors are not phase sensitive. The presence of the atomic medium maps phase noise into amplitude noise by various mechanisms~\cite{Cruz07,Zoller94}. This mapping leads to either correlation or anti-correlation between the fields and it is highly sensitive to any resonance in the sample. In the next subsections, it will be explained the details of the setups employed in the correlation measurements for each atomic species, performed at two different laboratories.

\subsection{Cesium Apparatus }
\label{sec:cesiummot}

\begin{figure}[ht]
\centering
\includegraphics[width=87mm]{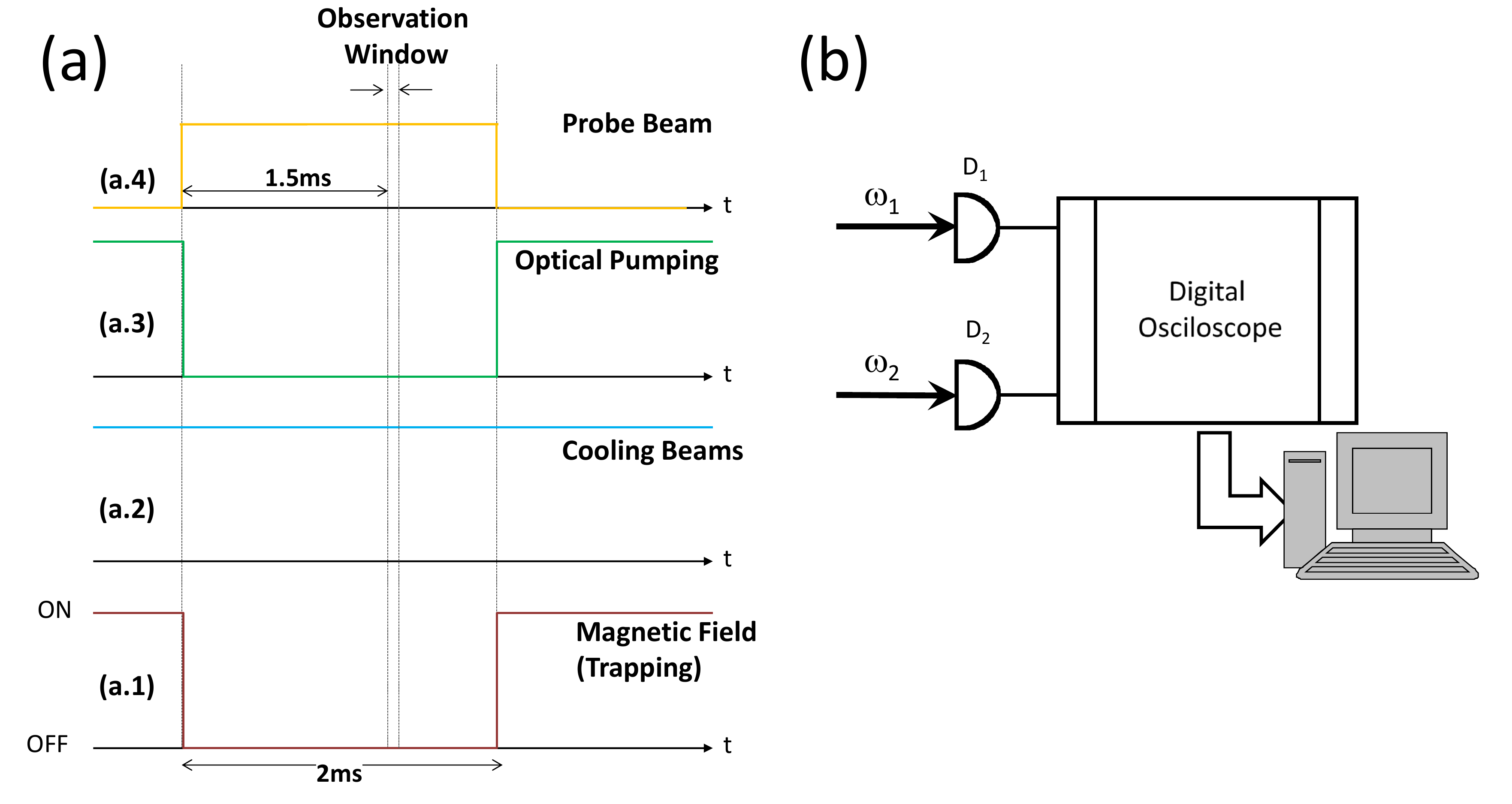}
\caption{ (a) Cesium pulse sequence for the noise spectroscopy measurement. On (a.1), (a.2), (a.3) we show the turn-off period for all trapping field, trap and repumper lasers and the quadrupole magnetic field. On (a.4) we show the period where the probe field is turned on and the position of the 8~$\mu$s observation window. (b) Detection scheme: After the interaction with the atomic medium, the beams are sent to two AC coupled photodiodes, $D_1$ and $D_2$. The photocurrents are registered by a digital oscilloscope for later analysis.}
\label{fig:csmot}
\end{figure}

In the cesium MOT~\cite{Tabosa06}, a glass chamber was employed that allows for  quick turning off of magnetic fields, typically in less than 1~ms~\cite{Moretti08}. A set of three pairs of coils in Helmholtz configuration maintains a close to null magnetic field in the region of the atomic cloud.  The final temperature of the atoms in the MOT is bellow 1 mK with an optical depht of 3 (around  $10^7$ atoms and a cloud diameter of 3 mm). The ECDL generating fields $1$ and $2$ is locked to the transition $6S_{1/2}(F=3)\rightarrow 6P_{3/2}(F'=2)$ and split in two parts, each beam passes through an independent AOM in double pass configuration, to minimize misalignments while tuning its frequenciy. The detected intensities are converted to photocurrents using two AC-coupled detectors (ET-2030A, Electro-Optics Technology) with a lower and upper cut off frequencies of 30~kHz and 1.2~GHz, respectively.  Measurements are performed in a 2~ms time window, in which the repump beam for the MOT (acting on the $6S_{1/2}(F=3)\rightarrow 6P_{3/2}(F'=3)$ transition) and the magnetic fields are turned off and cloud expansion can be neglected (Fig.~\ref{fig:csmot}). The photocurrents are recorded by a digital oscilloscope for 8~$\mu$s during each cycle, after the EIT fields act in the sample during 1.5~ms, giving time for a complete decay of stray magnetic fields and for the system to reach a steady state~\cite{Moretti08}.

We keep field $2$ detuning fixed on resonance, $\delta_2 = 0$, and tune the probe detuning $\delta_1$. For each $\delta_1$, 100 runs of the oscilloscope are recorded, which are later analysed by software. From the analysis, we obtain different measures of correlation between fields 1 and 2. The simplest of such quantities is the instantaneous, normalized cross-correlation function

\begin{equation}
g_2(0) = \frac{\langle \delta I_1(t) \delta I_2(t) \rangle}{\sqrt{\langle \delta I_1(t)^2 \rangle\langle \delta I_2(t)^2 \rangle}} \,,
\end{equation}

where $I_1(t)$ ($I_2(t)$) denotes the intensity of field $1$ ($2$) at time $t$, and $\langle \cdots \rangle$ denotes an average over time. It provides the most intuitive and direct approach to quantify such correlations. This approach was applied in a recent work to obtain a clear physical picture of some key spectral features observed in our measurements~\cite{Felinto13}. This time-based approach to measure and quantify correlations is also employed by other groups in most of recent works on the subject~\cite{Scully05,Scully10,Xiao09,Scully09}.

\begin{figure}[ht]
\centering
\includegraphics[width=80mm]{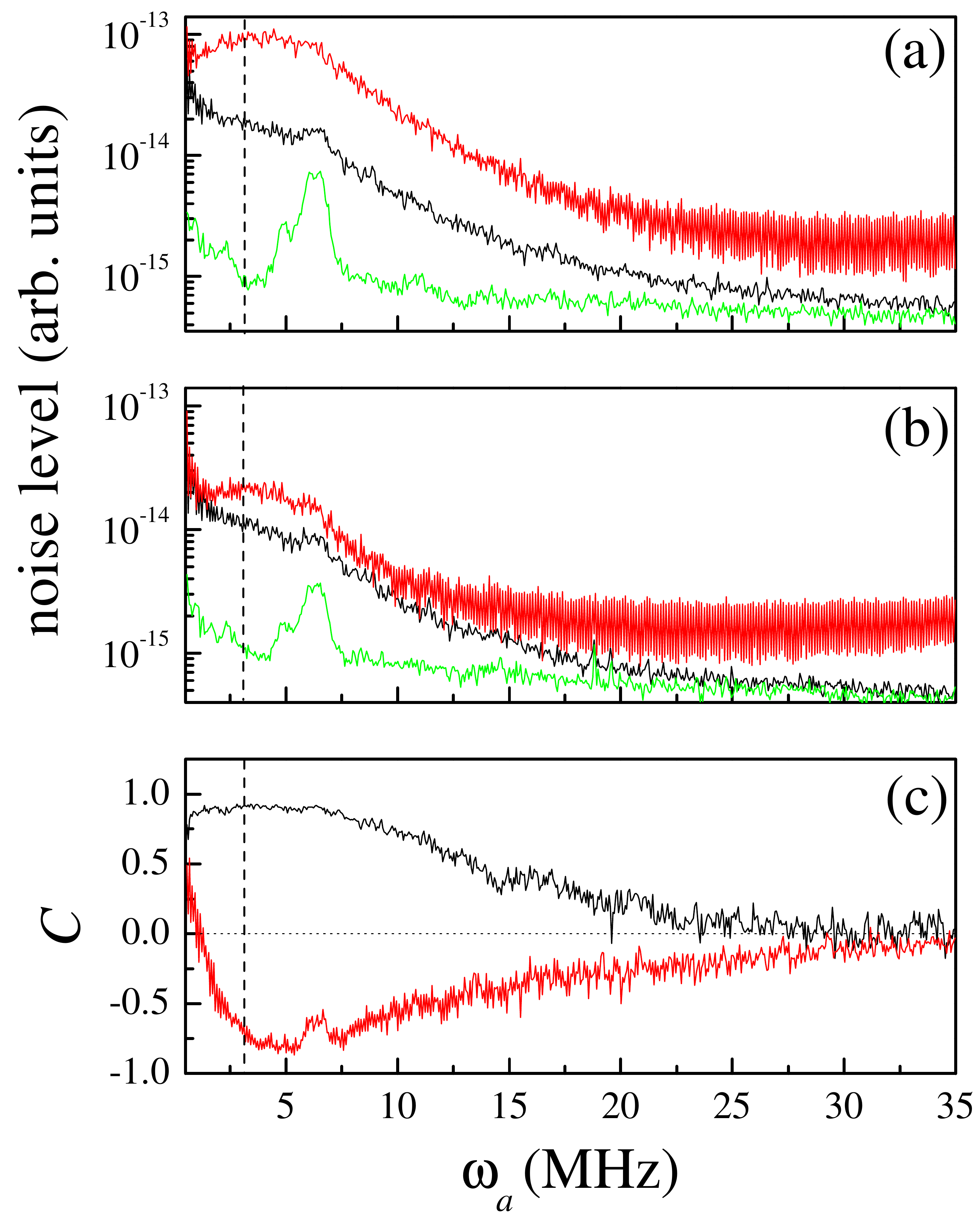}
\vspace{-0.5cm}
\caption{(a) and (b) plot the power spectra for the photocurrents associated to fields $2$ and $1$, respectively, for two different power levels: $P_2 = 30\,\mu$W (black) and~$500\,\mu$W (red), and $P_1 \approx P_2/2$, with $\delta_{2} = 0$ and $\delta_{1} = 0.73$ MHz. The respective dark electronic noise (green), obtained with the detectors blocked, is also plotted. (c) provides the correlation coefficient as a function of the analysis frequency for the data corresponding to the two power levels in (a) and (b). The dashed line indicates the analysis frequency employed to obtain an optimum correlation spectrum. The dotted line in (c) indicates the zero correlation level.}
\label{fig:fa}
\end{figure}

On the other hand, as an experimental method to acquire spectroscopic information on the atomic system, such temporal correlation functions are more susceptible to various sources of electronic noise that may decrease the observed degree of correlation, and consequently the signal-to-noise ratio of the spectral features. To clarify this point, we plot in Figs.~\ref{fig:fa}a and~\ref{fig:fa}b the noise spectra of the fields $2$ and $1$, respectively, for two different optical powers of field $2$, $P_2 \approx 30\,\mu$W and~$500\,\mu$W. The power in field $1$ is kept proportional to that on field $2$, such that $P_1 \approx P_2/2$. We also plot in each figure the corresponding ``dark'' electronic noise (fluctuations), measured by blocking all light going to the detector. We notice that there is a significant electronic noise in our system around 6~MHz. Also, as power increases, a broadband electronic noise affects our detectors, preventing them from reaching the dark electronic noise for high frequencies, where the contribution of the atomic system to the noise spectra should be negligible.     

A way to simplify problems like these is to switch to a spectral approach to measure and quantify the correlation between fields $1$ and $2$. This can be done by computing the normalized correlation coefficient $C$ for a particular frequency of analysis $\omega$~\cite{Cruz07}:
\begin{equation}
C(\omega) = \frac{S_{12} (\omega)}{\sqrt{S_{11}(\omega)S_{22}(\omega)} }
\label{eq:cor}
\end{equation}
where $S_{qq'}$ is the symmetrical correlation spectrum between the detected intensities,  with $q$ and $q'$ indicating fields 1 or 2, calculated as
\begin{eqnarray}
S_{qq'}(\omega) &=& \int^{\infty}_{-\infty}[\langle I_{q}(t+\tau)I_{q'}(t)\rangle\nonumber \\ 
&-& \langle I_{q}(t+\tau)\rangle \langle I_{q'}(t)\rangle]\times e^{i \omega \tau}d\tau \,.
\label{eq:sqq}
\end{eqnarray}
$C(\omega)$ is normalized such that $C(\omega)=1$ for maximum correlation and $C(\omega)= -1$ for maximum anti-correlation. $C(\omega)=0$ indicates completely uncorrelated fluctuations. 

\begin{figure}[ht]
\centering
\includegraphics[width=80mm]{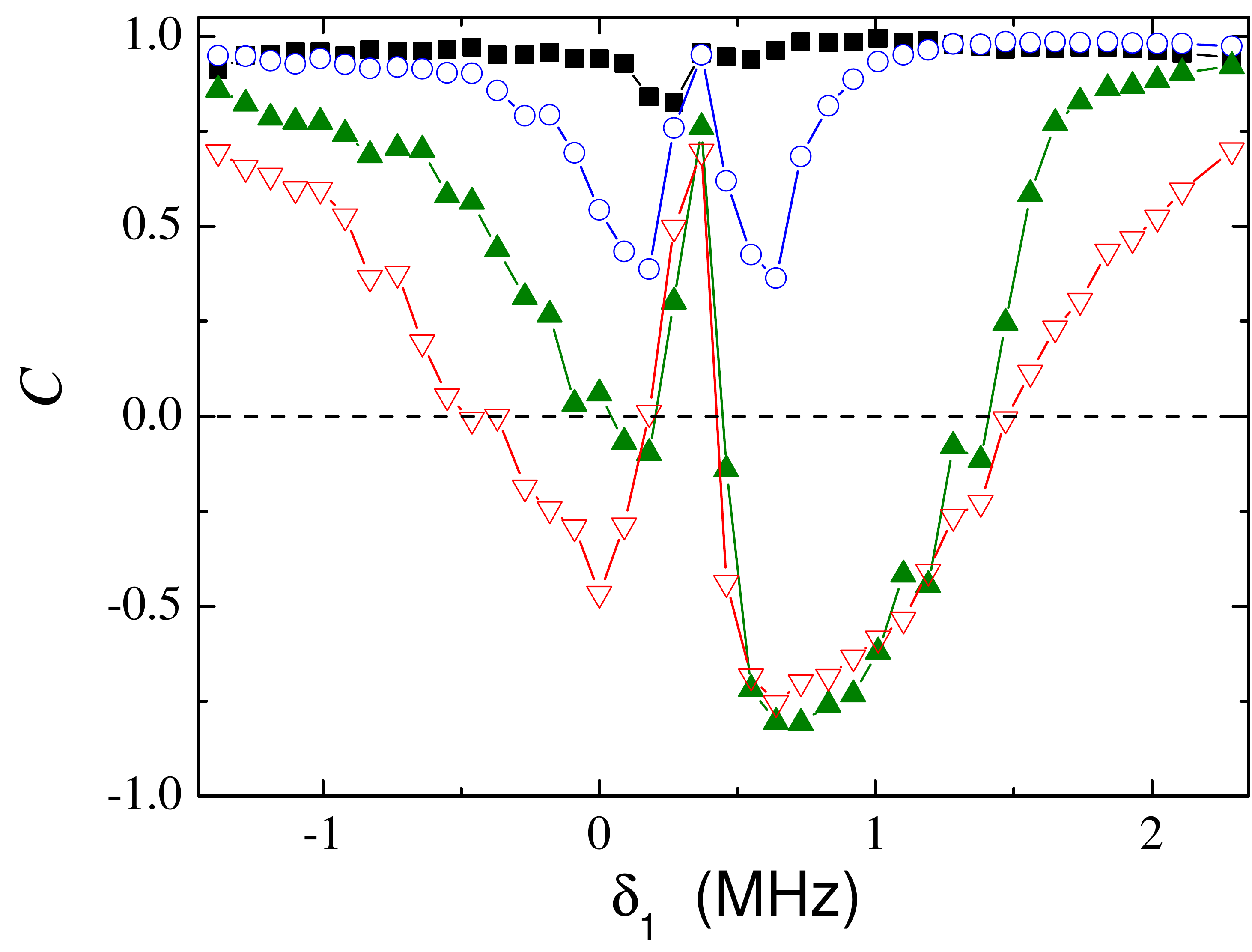}
\caption{Correlation coefficient as a function of the detuning of field $1$ for various power levels. From top to bottom, we have $P_1 = 30\,\mu$W (black squares), 100$\,\mu$W (empty blue circles), 300$\,\mu$W (green triangles), and 500$\,\mu$W (empty red triangles), respectively, with $P_1 \approx P_2/2$. The analysis frequency is fixed at $\omega_a = 3\,$MHz. The lines are just guides for the eyes.}
\label{fig:cscorr}
\end{figure}

In Fig.~\ref{fig:fa}c we plot the spectral correlation function from the signals at Figs.~\ref{fig:fa}a and~\ref{fig:fa}b. As expected, the strongest correlations appear in the region around 3~MHz, where the atomic system introduces the largest noise components, giving the highest ratio of correlated noise to uncorrelated electronic fluctuations. The time correlations would average all these contributions for different analysis frequency, resulting in smaller degrees of correlation. On the other hand, by selecting a particular analysis frequency in the spectral region presenting the highest degrees of correlation, we enhance any spectral feature appearing as $\delta_1$ is tuned. The results for this approach for generating a correlation spectrum is presented in Fig.~\ref{fig:cscorr}, in which we plot the correlation coefficient $C$ as a function of $\delta_1$ for a fixed $\omega_a = 3\,$MHz and various power levels of the exciting fields.

Figure~\ref{fig:cscorr} shows the characteristic behavior of the EIT correlation spectra in terms of the detuning $\delta_{1}$ as the power level changes~\cite{Felinto13}. For low powers, the system presents only correlations. As power increases, correlation decreases on the sides of the two-photon, EIT resonance, until it turns into anti-correlations for high enough powers. The result is a narrow correlation peak that is power-broadening independent for high powers, whose linewidth is directly related to the ground-state-coherence lifetime~\cite{Xiao09,Felinto13}. The shift of the correlation peak from $\delta_{1} = 0$ is due to a residual DC magnetic fields. The comparison of the results of Fig.~\ref{fig:cscorr} with the corresponding $g_2(0)$ plot in Ref.~\onlinecite{Felinto13} illustrates the clear advantage of the present approach, with stronger correlations and anti-correlations maximizing the contrast of the curve around the correlation peak.

\subsection{Rubidium Apparatus}
\label{sec:rubidiummot}

The frequency-domain approach to measure the correlation properties of the fields participating in the EIT process was fully explored in a second series of measurements conducted in a rubidium MOT. In this case, the frequency of analysis was chosen first, and then the detection system was optimized accordingly. For each detector, the DC and AC part of the signal at this particular frequency of analysis was acquired and sent to a computer, which can process the information and calculate the correlation coefficient $C$ immediately. The correlation spectrum was plotted in real time as the detuning $\delta_1$ is scanned, opening the possibility of using such spectrum for real-time optimizations of the system.

\begin{figure}[htb!]
\centering
\includegraphics[height=47mm,width=86mm]{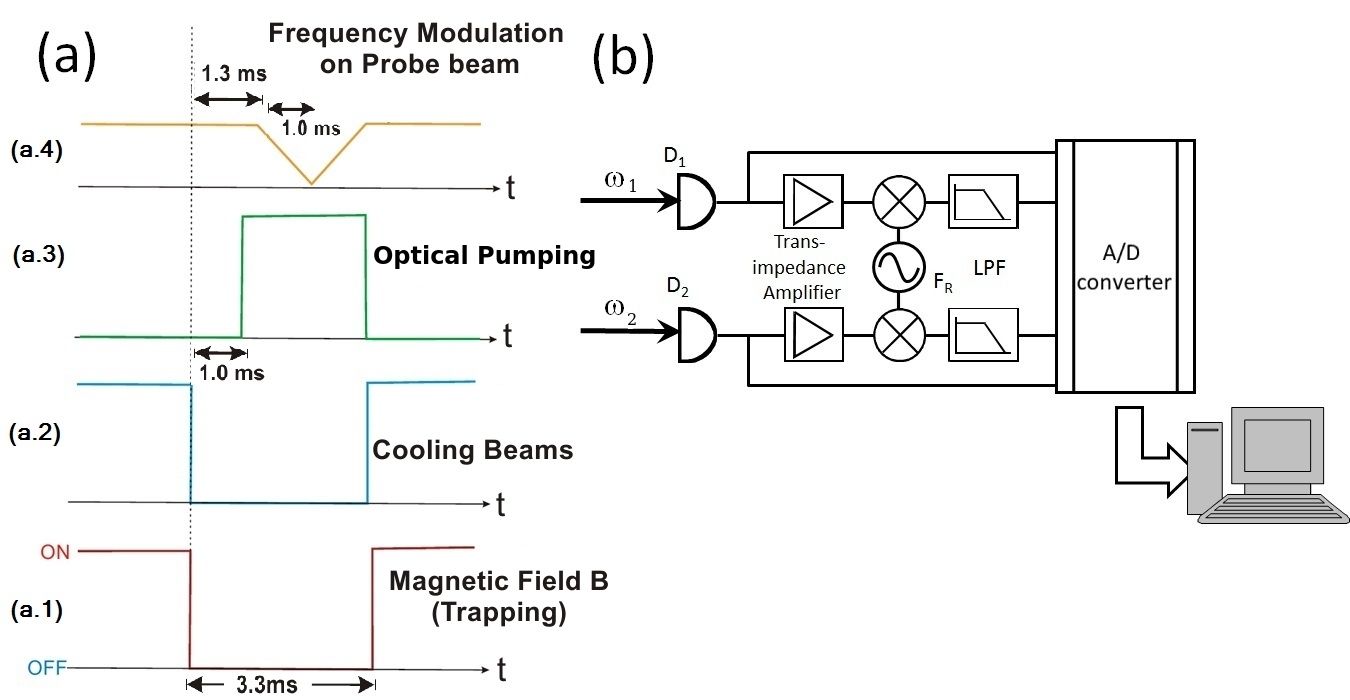}
\caption{(a) Time control sequence of the parameters for the noise spectroscopy measurements. On  (a.1) and (a.2), the MOT quadrupole magnetic field and cooling beams are switched-off. On (a.3) we switch-on the optical pumping field 1~ms after the magnetic field was switched-off. On (a.4), the frequency modulation is performed during 1~ms. (b) Detection scheme: After interaction with the atomic medium, the beams were sent to detectors that separate the high frequency signal (AC signal) from its low frequency component (DC signal). The high frequency signal is demodulated  (the analysis frequency is set up in the mixer) and the data is digitalized and stored.}
\label{fig:rbmot}
\end{figure}

The Rubidium MOT has a setup similar to the apparatus for Cesium presented in the previous section. This system traps $^{85}$Rb isotope. The atomic cloud has around $ 7 \times 10^{7}$ atoms and a diameter of $4$ mm. The average atomic temperature is $205 \mu K$. The $1$ and $2$ beams are obtained from a single ECDL, independent from the MOT beams, locked to a crossover of the $F=2 \rightarrow F'$ transition of the D2 line of Rubidium and frequency shifted to the $5S_{1/2}(F=2)\leftrightarrow 5P_{3/2}(F'=2)$ resonance. The timing of each realization of the experiment is shown in Fig.~\ref{fig:rbmot}(a). The magnetic field and the cooling beams of the MOT (trapping and repump) are turned off for 3.3~ms. After these fields are off for 1~ms, an optical pumping beam is turned on, close to resonance for the $5S_{1/2}(F=3)\leftrightarrow 5P_{3/2}(F'=3)$ transition. This beam is kept on for the remaining period when the MOT is off, pumping all atoms to the ground state $5S_{1/2}(F=2)$. The probe beam is kept on all the time. After the optical pumping acts on the system for 300~$\mu$s, a linear scan of the probe-beam frequency starts, continuing for 1~ms. In this way, differently from the Cesium experiment, the whole correlation spectrum is measured at each realization of the experiment.
 
As pointed out above, the detection system has also different features from the previous experiment, as shown in Fig. \ref{fig:rbmot}(b). Here the photocurrent of the PIN photodiodes is splited between a DC amplifier and a HF trans-impedance amplifier. The DC part of the signal is sent to an  analog-to-digital (AD) converter. The AC component is demodulated at a given analysis frequency by mixing with an electronic oscillator and low-pass filtering the output. The electronic oscillator frequency was set to 2 MHz, and the low-pass filter has a bandwidth of 300 kHz. As a result, the AD converter receives the amplitude of the noise spectra, measured at the oscillator frequency with a bandwidth of 600 kHz. These signals from the two detectors can be combined in order to calculate the correlation coefficient $C$ while scanning $\delta_1$. The curves from various scans are averaged in order to obtain the final result.  

\begin{figure}[hbt!]
\centering
\vspace{0.7cm}\includegraphics[height=67mm,width=55mm]{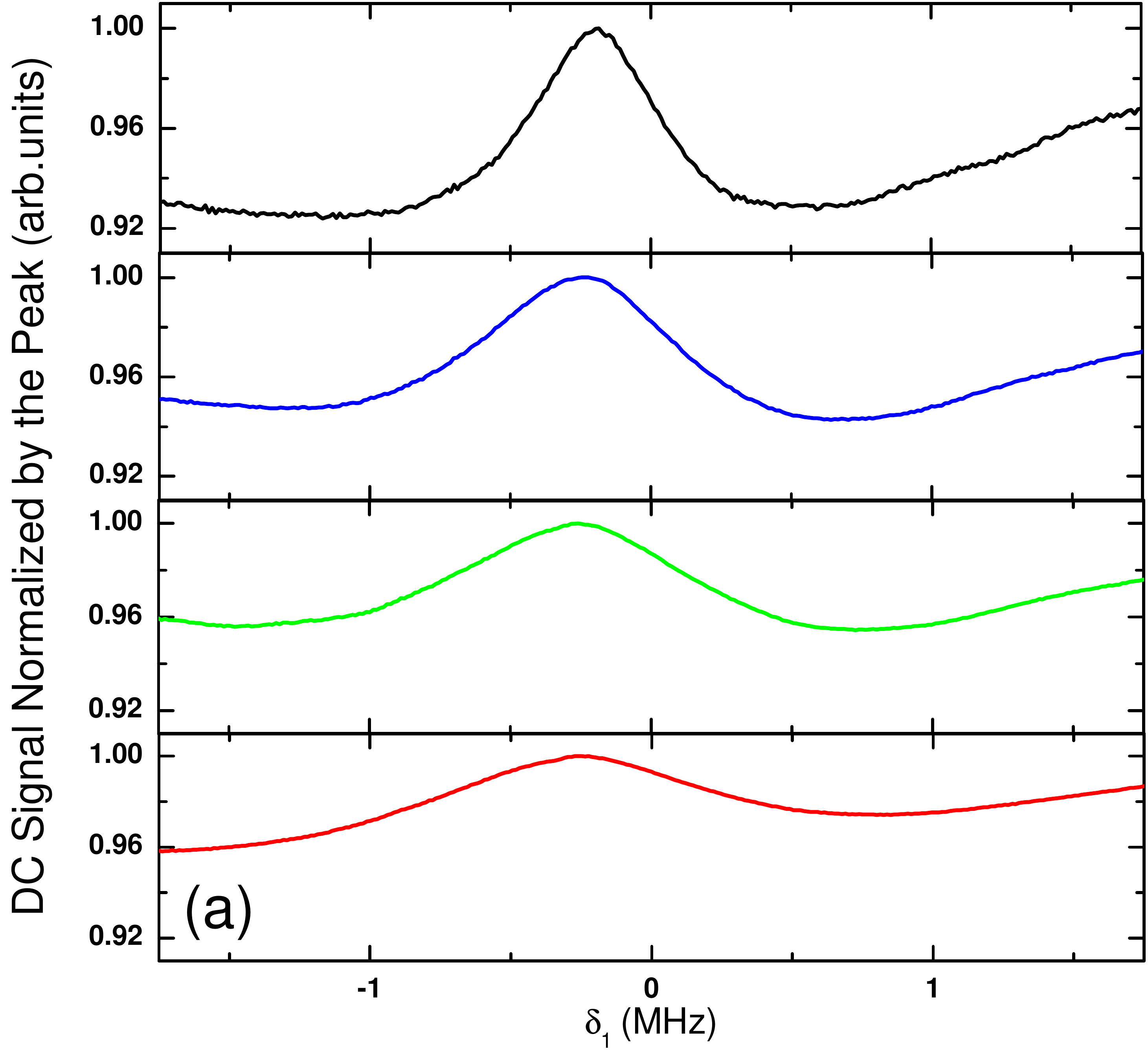}\hspace{0.2cm}\includegraphics[height=67mm,width=55mm]{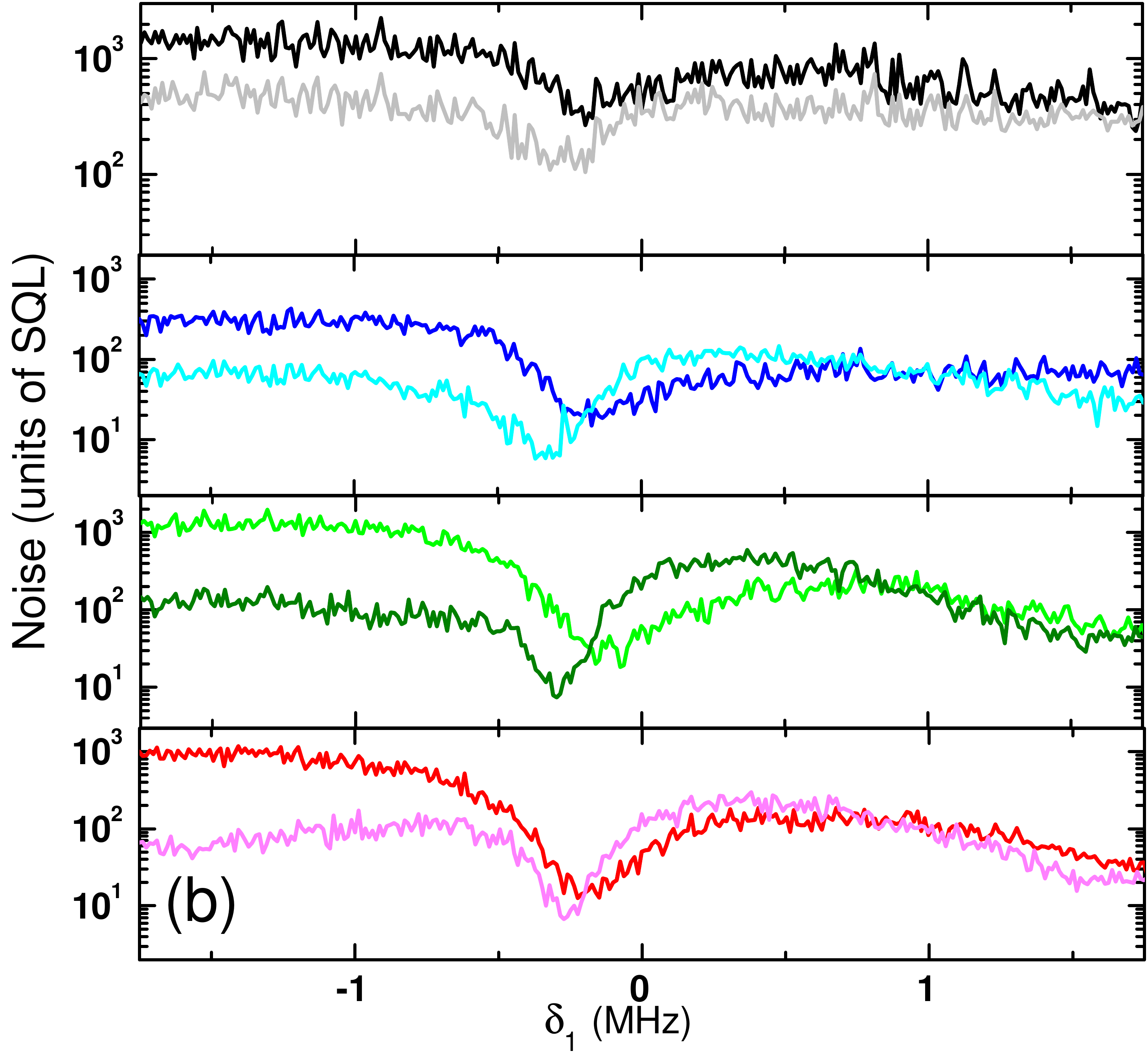}\hspace{0.2cm}\includegraphics[height=67mm,width=55mm]{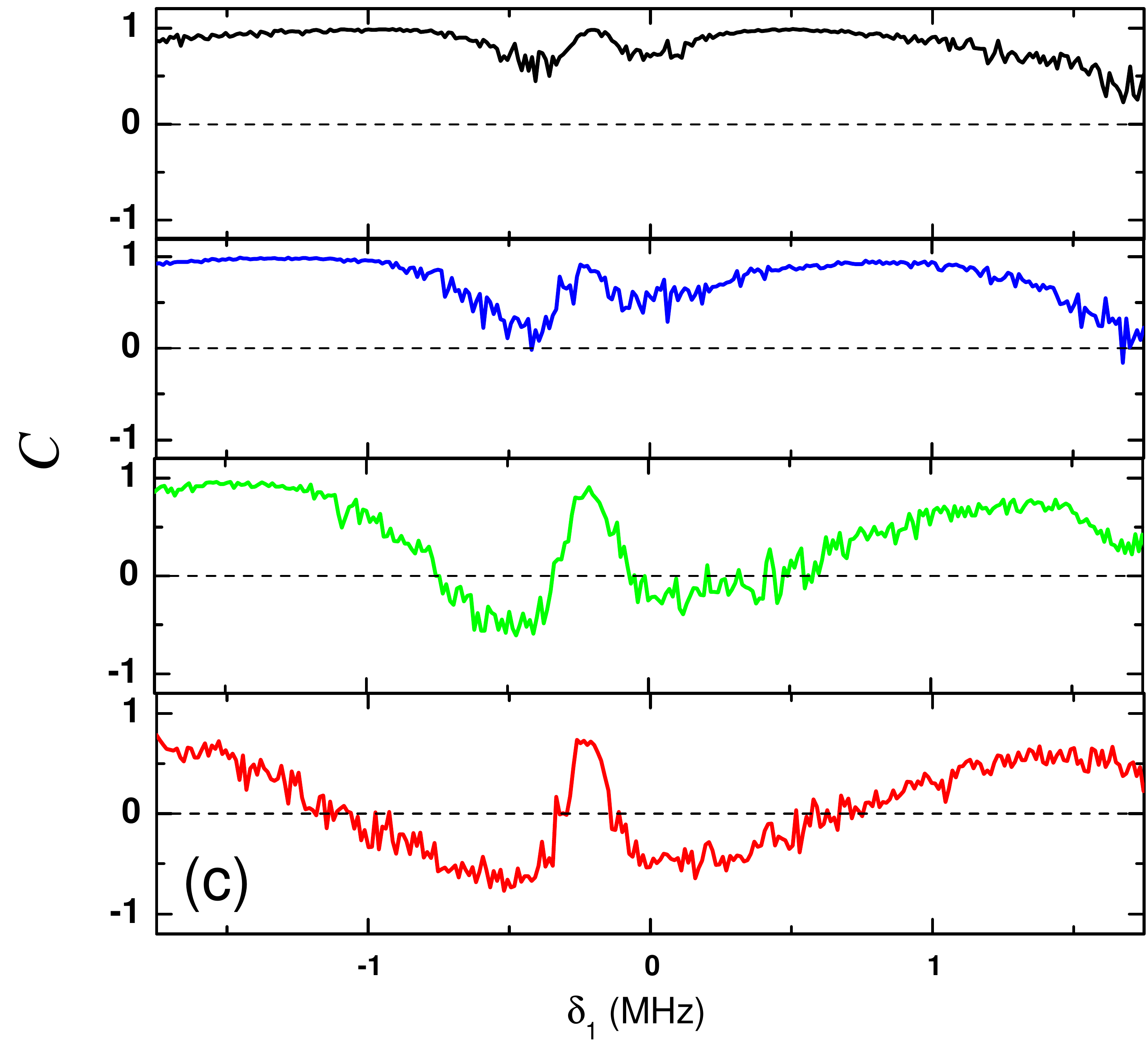}
\caption{(a) DC signal, (b) noise of fields $1$ and $2$, and (c) correlation coefficient versus detuning of field $1$ (the frequency of field $2$ is kept in resonance) for different powers of the optical fields. For (a), (b) and (c) from top to bottom the powers are 30~$\mu$W (black), 100~$\mu$W (blue), 200~$\mu$W (green), and 300~$\mu$W (red). For (b) from top to bottom (grey), (cyan), (dark green) and (pink) are the curves for the beam 2. Data obtained from the Rubidium MOT.}
\label{fig:rbdcHFCorr_signal}
\end{figure}

Figure~\ref{fig:rbdcHFCorr_signal} shows then the data for four different laser intensities, corresponding to powers of 30~$\mu$W, 100~$\mu$W, 200~$\mu$W, and 300~$\mu$W, respectively.  In Fig.~\ref{fig:rbdcHFCorr_signal}(a), we have the usual EIT peak, appearing on the DC signal once the two-photon resonance condition is satisfied. Figure~\ref{fig:rbdcHFCorr_signal}(b) presents the noise power of each beam separately as a function of $\delta_1$.  Figure~\ref{fig:rbdcHFCorr_signal}(c) displays the corresponding correlation spectra. The qualitative behavior in Fig.~\ref{fig:rbdcHFCorr_signal}(c) is very similar to the one in Fig.~\ref{fig:cscorr}. The noise power in Fig.~\ref{fig:rbdcHFCorr_signal}(b) shows that the narrow peak in the correlation signal does not appear in the noise signal of the individual beams.  

\begin{figure}[ht]
\centering
\includegraphics[width=67mm]{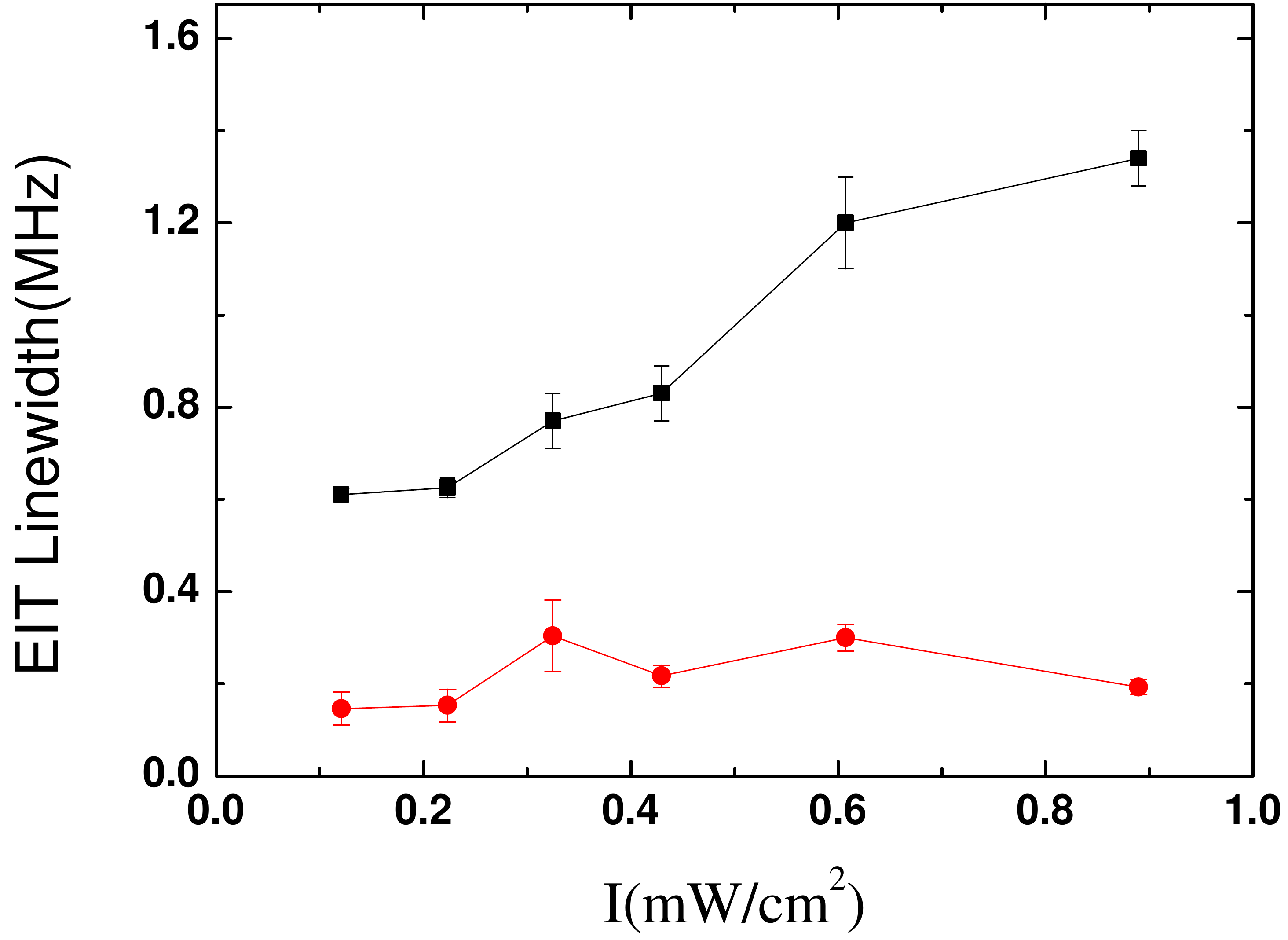}
\caption{Linewidth measured through the DC signal (squares) and the correlation spectrum (circles) as a function of excitation power. Data obtained for the Rubidium MOT. The lines are just guides for the eyes.} 
\label{fig:experbroad}
\end{figure}

The comparison between Figs.~\ref{fig:rbdcHFCorr_signal}(a) and (c) provides also a direct example of the advantage of such correlation spectroscopy over the usual technique to measure the EIT spectrum on the DC signal. Figure~\ref{fig:rbdcHFCorr_signal}(a) shows the broadening of the EIT peak as power increases, while in Fig.~\ref{fig:rbdcHFCorr_signal}(c) the same peak in the correlation spectrum has a power-broadening-independent linewidth. This is an important advantage of this technique, which was already highlighted in the previous section and explored in previous references~\cite{Xiao09,Felinto13}. The EIT peak for the lowest power was already broadened when compared to the corresponding structure in the correlation spectrum. All these observations are summarized in the plots of Fig.~\ref{fig:experbroad}, providing the linewidth of the EIT peak for both DC signal and correlation spectrum as a function of power of the excitation fields.

\onecolumngrid

\twocolumngrid

\section{Theoretical modeling}
\label{sec:theory}

A simplified theory in time domain explaining spectral structures like the ones in the previous section was introduced in Ref.~\cite{Felinto13}. The physical picture emerging from that model reveals that the Doppler-broadening-free correlation peak at the EIT resonance is a result of the competition between the absorptive and dispersive parts of the atomic coherence, with both contributing to the transmitted noise of the fields due to the mechanism of conversion of phase-noise to amplitude-noise by the atomic medium~\cite{Felinto13}. Such theory was able to model well the same time correlations observed in the fields, but its heuristic approach can not be extended to the frequency-domain analysis introduced in the present work. In order to model the spectra presented in the previous section for a particular analysis frequency, we introduce a more standard theory for the correlation spectra based on Ito calculus, following the approach of Ref.~\cite{Zoller94} to model the interaction of a generic atomic system with lasers presenting only phase noise. Previously, such approach has been already successfully applied to model correlation spectra in vapor cells at room temperature~\cite{Martinelli04,Cruz07}.  

\begin{figure}[ht]
\centering
\includegraphics[width=50mm]{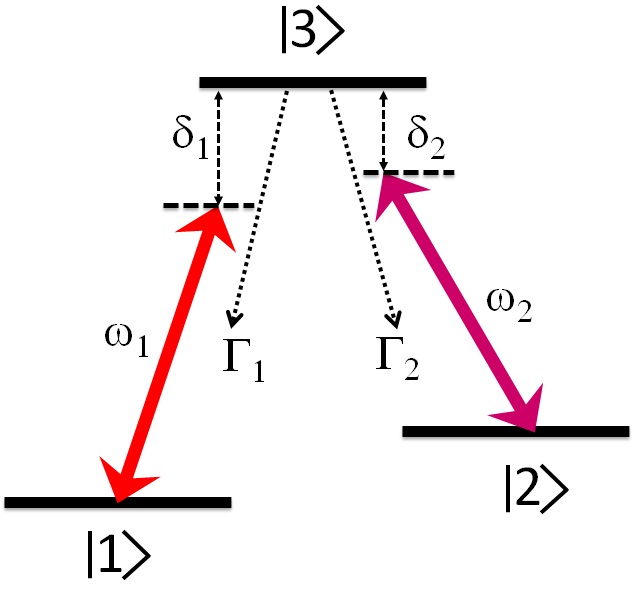}
\caption{Theoretical model: three-level system in Lambda  ($\Lambda$) configuration. The solid lines represent the two electromagnetic fields interacting with the atoms. The detuning of field $i$, with frequency $\omega_i$, from the excited state is given by $\delta_i$. $\Gamma_1$ ($\Gamma_2$) is the excited state decay rate to level $|1\rangle$ ($|2\rangle$).}
\label{fig:3Nsis}
\end{figure}

In order to model all key features of our experimental results, it is sufficient to consider a three-level atom interacting with two laser fields in $\Lambda$ configuration (see Fig.~\ref{fig:3Nsis}). It is well known that such simple system explains well many aspects of degenerate two-level systems, once the hiperfine quantum number $F_G$ for the ground state is equal or larger than the one $F_E$ for the excited state~\cite{Arturo99}. We label the two ground states as levels $|1\rangle$ and $|2\rangle$, and the excited state as level $|3\rangle$. The transition frequencies between levels $|1\rangle$ ($|2\rangle$) and $|3\rangle$ is $\omega_{13}$ ($\omega_{23}$). $\Gamma_1$ ($\Gamma_2$) is the excited  state, natural decay rate to level $|1\rangle$ ($|2\rangle$). $\omega_{i}$ is the frequency of field $i$. 

We assume that the atoms are excited by two electromagnetic fields with constant amplitudes and orthogonal polarizations, but some stochastic phase noise, representing the typical condition of diode lasers as those used in the experiments~\cite{Zhang95}:

\begin{equation}
\mathbf{E}_{i}(t)={\cal E}_{i} \exp \left[i(\omega_{i} t+ \phi_{i})\right] \mathbf{e}_{i} \,,
\label{eq:field}
\end{equation}
where the index $i =1,2$ labels the field addressing the transitions with ground states $|1\rangle$,$|2\rangle$, respectively.  The complex amplitudes of the fields are given by ${\cal E}_{i}$, and the frequencies and polarizations  of the lasers are $\omega_{i}$ and $\mathbf{e}_{i}$, respectively.  The time evolution of the phases $\phi_{i}$ are represented by  two stochastic variables describing Wiener processes \cite{Gardiner-book}, satisfying the relations: 
\begin{eqnarray}
\langle d\phi^2_{1} \rangle& =&\langle d\phi^2_{2} \rangle = \langle d\phi_{1} d\phi_{2}\rangle = 2bdt \,, \nonumber\\
\langle d\phi^n_{1} \rangle &=&\langle d\phi^n_{2} \rangle=0, \;(n \neq 2) 
\label{eq:devphi}
\end{eqnarray}

with  $2b$ representing the laser full linewidth. We also assumed above that the two fields come from the same laser, so that their fluctuations are perfectly correlated. 
 As anticipated above, this phase noise is the only source of noise in our model, being in the origin of all amplitude noise in the output signals.  

The  Hamiltonian that describes the interaction between atoms and fields is given by: 

\begin{eqnarray}
H(t) &=&  \hbar\omega_{13} | 1\rangle\langle 1| + \hbar\omega_{23}  | 2\rangle\langle 2|] + [\hbar \Omega_{1} e^{ i(\omega_{1} t+ \phi_{1}) }| 1\rangle\langle 3| \nonumber\\
 &+&\hbar \Omega_{2}e^ {i(\omega_{2} t+ \phi_{2})} | 2\rangle\langle 3| +h.c.] 
 \label{eq:hamilton}
\end{eqnarray}
where the Rabi frequencies for the electromagnetic fields,  $\Omega _1=-\frac{\mu _{13}{\cal E}_1}\hbar $ and $\Omega _2=-\frac{\mu_{23}{\cal E}_2}\hbar $, provide the strength of the coupling between fundamental and excited states. The terms $\mu_{13}$ and   $\mu_{23}$ are the atomic dipole moments associated with the transitions $ |1\rangle \rightarrow  |3\rangle$ and $|2\rangle \rightarrow  |3\rangle$, respectively. The Bloch equations for the density matrix describing the state of the system are then: 
\begin{eqnarray}
\dot{\rho}_{11} &=& \left[-i\Omega_{1}^{*}e^{-i(\omega_{1} t +\phi_{1})}\rho_{13}+ c.c.\right] + \frac{\Gamma}{2}\left( 1-\rho_{11}-\rho_{22}\right) \,,\nonumber\\
\dot{\rho}_{22} &=& \left[-i\Omega^{*}_{2}e^{-i(\omega_{2} t +\phi_{2})}\rho_{23}+c.c.\right]  + \frac{\Gamma}{2}\left( 1-\rho_{11}-\rho_{22}\right) \,, \nonumber\\ 
\dot{\rho}_{13} &=& i\omega_{13}\rho_{13}+i\Omega_{1}e^{i(\omega_{1} t +\phi_{1})}(1-\rho_{22}-2\rho_{11}) \nonumber\\
&& -i\Omega_{2}e^{(i\omega_{2}t +\phi_{2})}\rho_{12}-\frac{\Gamma}{2}\rho_{13}\,,\nonumber\\
\dot{\rho}_{23} &=& i\omega_{23}\rho_{23}+i\Omega_{2}e^{i(\omega_{2} t +\phi_{2})}(1-\rho_{11}-2\rho_{22})\nonumber\\
 && -i\Omega_{1}e^{(i\omega_{1}t +\phi_{1})}\rho_{21}-\frac{\Gamma}{2}\rho_{23}\,,\nonumber\\
\dot{\rho}_{12} &=& i(\omega_{13}-\omega_{23})\rho_{12}+i\Omega_{1}e^{i(\omega_{1} t +\phi_{1})}\rho_{32}\nonumber\\
&-&i\Omega^{*}_{2}e^{i(\omega_{2} t +\phi_{2}()}\rho_{13} -\gamma\rho_{12}\,,
\label{eq:bloch_3N}
\end{eqnarray}	

where we assumed $\Gamma_1 = \Gamma_2 = \Gamma/2$. Decoherence between the two ground states is introduced through an effective homogeneous decay rate $\gamma$\cite{Moretti08,deOliveira12}.

In the limit of a thin sample, the output field after interaction with the atoms can be expressed as~\cite{Zoller94}
\begin{equation}
\mathbf{E}_{out}(t) = \mathbf{E}_1(t) +\mathbf{E}_2(t) + i \frac{\beta}{2c\epsilon_{0}}\mathbf{P}(t)
\label{eq:output}
\end{equation}

where the atomic response to the initial fields $\mathbf{E}_i$ 
is given by the complex polarization $\mathbf{P}(t)$ of the medium.  The parameter $\beta$ is a real constant proportional to the medium's optical density. The polarization can be write explicitly in terms of the components for each transition 
\begin{eqnarray}
\mathbf{P}(t) = \mathbf{p}_1(t) +  \mathbf{p}_2(t) \,,
\label{eq:p}
\end{eqnarray}
with $p_{1} \propto \rho_{13}$ and $p_{2} \propto \rho_{23}$. The output field including the polarization contribution is used to evaluate the detected intensity defined by $I_{q}(t) =2  c \epsilon_{0} | \mathbf{E}_{out}(t).\mathbf{e}_{q}(t)|^2$, where $q=1,2$. Finally, the power spectra $S_{qq^{\prime}}(\omega)$ at frequency $\omega$ can be obtained from Eq.~(\ref{eq:sqq}). The correlation functions in Eq.~(\ref{eq:sqq}) are calculated from the Bloch equations~(\ref{eq:bloch_3N}) following directly the procedure of Ref.~\cite{Cruz07}, neglecting the Doppler broadening of the sample.
 
\begin{figure}[htb!]
\centering
\includegraphics[width=80mm]{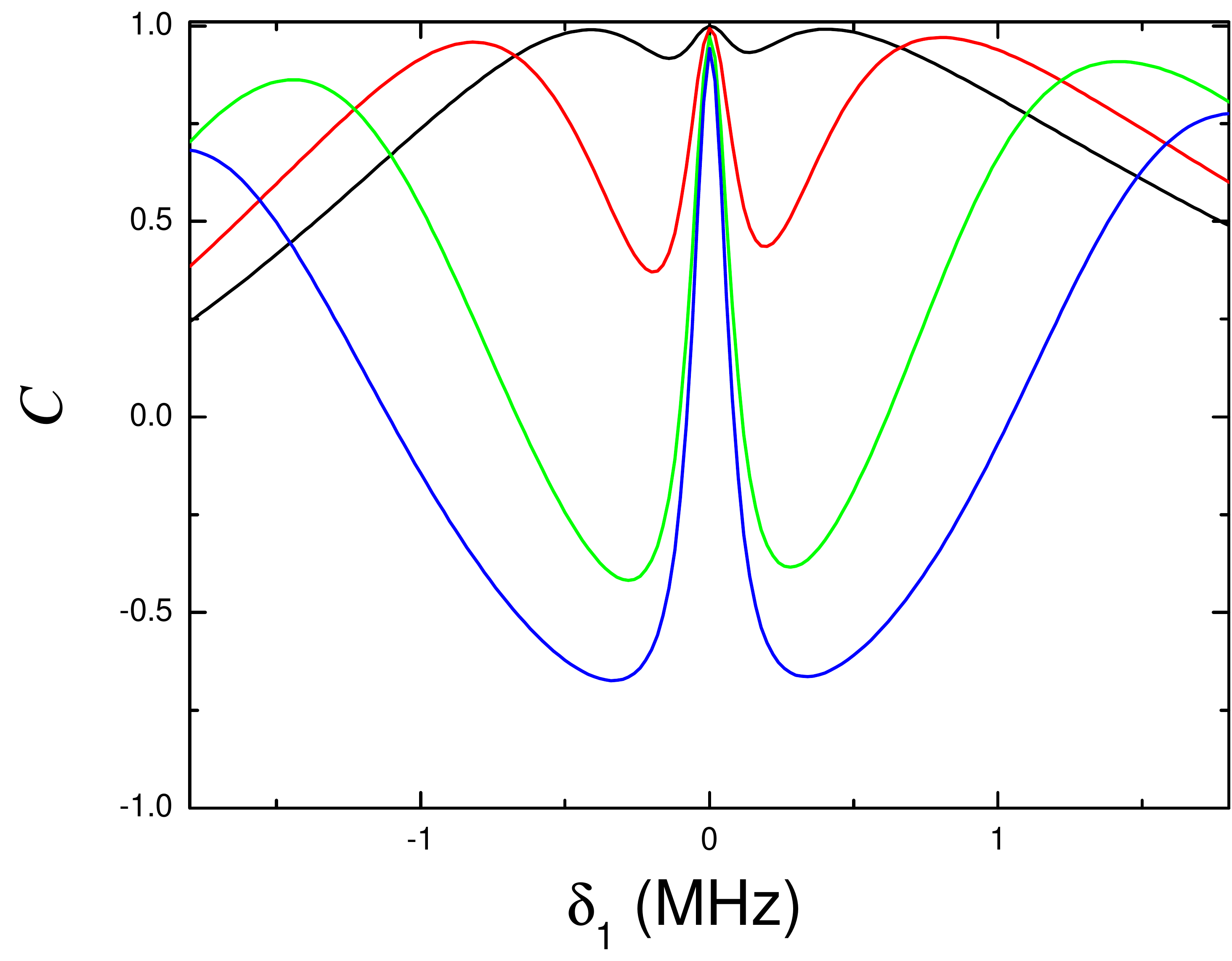}
\caption{(a) Numerical calculation of correlation coefficient versus detuning of field 1 for different Rabi frequencies, with $\Omega_1=0.75 \Omega_2$ and (black) $\Omega_2 = 0.09\Gamma$, (red) $\Omega_2 = 0.17\Gamma$, (green) $\Omega_2 = 0.29\Gamma$, and (blue) $\Omega_2 = 0.37\Gamma$.  Other relevant parameters are $\delta_2 = 0$, $b/2\pi =1$ MHz, $\omega_a/2\pi =3$ MHz, $\gamma/2\pi =150$ kHz, and $\Gamma/2\pi = 5.2$~MHz. The calculation considered a relation between Rabi frequencies consistent to the relation between intensities for the experimental data of Fig.~\ref{fig:cscorr} for cesium atoms.} 
\label{fig:teorabi}
\end{figure}

\begin{figure}[ht]
\centering
\includegraphics[width=80mm]{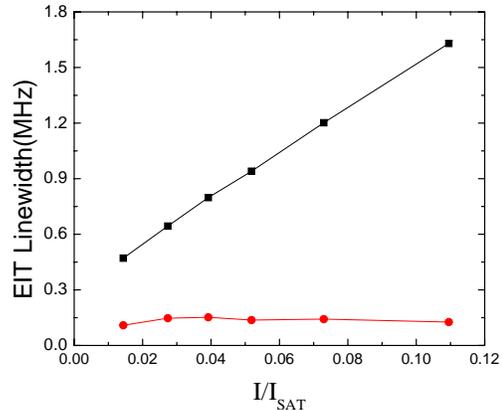}
\caption{Numerical calculation of the EIT linewidth on (black squares) DC signal and (red circles) correlation coefficient versus Rabi frequency. In these calculations, we used $\Omega_1=\Omega_2$, $\delta_2=0$,  $b/2\pi =1$MHz, $\omega_a/2\pi =2$MHz, $\gamma/2\pi =150$kHz, and $\Gamma/2\pi = 6$~MHz. The calculation considered a situation close to the experimental data of Fig.~\ref{fig:experbroad} for rubidium atoms.} 
\label{fig:teobroad}
\end{figure}

The results of the model for the correlation coefficient as a function of detuning of one of the fields from the two-photon resonance condition are shown in Fig.~\ref{fig:teorabi}. The parameters in this figures were adjusted to approximate the experimental situation in Fig.~\ref{fig:cscorr} for cesium atoms, even though the same behavior is observed for the rubidium data in Fig.~\ref{fig:rbdcHFCorr_signal}. The model predicts quite well most of the observed features, confirming that the above theoretical approximations, including the assumption of three-level atoms, were enough to capture the essential aspects of the problem. An interesting difference between theory and experiment is that the correlations are typically higher in the experimental data. We understand this difference as coming from our assumption that we have a thin sample, since a thicker sample could in principle induce even more correlation. Another difference is the asymmetry of the spectrum around the central correlation peak. This could not be reproduced by our simple theory, and it might come from different optical pumping conditions over the atomic Zeeman structure as the detuning is changed. The dislocation from zero detuning of the correlation peak can be easily understood as coming from a residual magnetic field in the direction of propagation of fields 1 and 2, which dislocates the two-photon resonance uniformly over the whole ground-state Zeeman structure.

Another relevant feature in the experimental correlation spectra of last section is that the central correlation peak width remains unchanged over a wide range of Rabi frequencies. This is clearly revealed in the comparison between DC-signal and correlation-coefficient spectra in Fig.~\ref{fig:rbdcHFCorr_signal}, which presents significant broadening for the DC signal and no power broadening for the correlation coefficient. The same behavior is observed in the theoretical data of Fig.~\ref{fig:teobroad}, which was plotted in a situation close to the conditions of Fig.~\ref{fig:rbdcHFCorr_signal} for the rubidium experiments.

\section{conclusions}
\label{sec:conclusions}



In this work we measured the noise correlation spectrum using two different samples of cold atoms in EIT regime. The correlation spectroscopy was done in two different systems using different techniques. The use of cold atoms allowed the direct access to a specific EIT resonance, and we could discard the effect of other transitions, present in Doppler-broadened samples.

We have successfully shown that correlation spectroscopy provides a direct measurement of the coherence linewidth in this EIT transition, being insensitive to power broadening in the studied range of intensities. The simple model we used, relying in a three level system, was sufficient to describe the main features observed in more complex Zeeman manyfold EIT structures, for different atomic media. Although the richness of the involved structures limits the model to a qualitative description of the problem, in a 3-level $\Lambda$ system it should provide a quantitative agreement with the experiment.

 Moreover, the use of spectral analysis in the detection process provided a clearer signal than the temporal approach previously used, and a real-time reading of the coherence lifetime.  This is a crucial step for developing metrological applications from this technique. It can also be applied for the optimization of the coherence time for atomic memories used in quantum information protocols. 

This work was supported by grant \#2010/08448-2, S\~ao Paulo Research Foundation (FAPESP),  CNPq, CAPES and FACEPE (Brazilian agencies), through the programs PROCAD, PRONEX, and INCT-IQ (Instituto Nacional de Ci\^encia e Tecnologia de Informa{\c c}\~ao Qu\^antica). The authors would like to thank P. Nussenzveig for helpful discussions.  M. Martinelli would like to thank J. G. Aguirre G\'{o}mez for fruitful discussions.

\end{document}